\documentclass[conference]{IEEEtran}
\IEEEoverridecommandlockouts
\usepackage{cite}
\usepackage{amsmath,amssymb,amsfonts}
\usepackage{algorithmic}
\usepackage{graphicx}
\usepackage{textcomp}
\usepackage{xcolor}
\usepackage{subfigure}
\def\BibTeX{{\rm B\kern-.05em{\sc i\kern-.025em b}\kern-.08em
    T\kern-.1667em\lower.7ex\hbox{E}\kern-.125emX}}
\begin{document}

\title{Blind Estimation of Sub-band Acoustic Parameters from Ambisonics Recordings using Spectro-Spatial Covariance Features\\
\thanks{The work for this paper was conducted as a Research
Internship at Dolby Laboratories Australia.}
}
\author{
\begin{tabular}{@{}c@{}}
Hanyu Meng\textsuperscript{*,†},
Jeroen Breebaart\textsuperscript{†},
Jeremy Stoddard\textsuperscript{†},
Vidhyasaharan Sethu\textsuperscript{*},
Eliathamby Ambikairajah\textsuperscript{*}
\end{tabular}
\\
\IEEEauthorblockA{\textit{University of New South Wales\textsuperscript{*}, Dolby Laboratories, Australia\textsuperscript{†}}}
\IEEEauthorblockA{\{hanyu.meng,e.ambikairajah,v.sethu\} @unsw.edu.au}
\IEEEauthorblockA{\{jeroen.breebaart,jeremy.stoddard\} @dolby.com}
}

\maketitle

\begin{abstract}
Estimating frequency-varying acoustic parameters is essential for enhancing immersive perception in realistic spatial audio creation. In this paper, we propose a unified framework that blindly estimates reverberation time (T60), direct-to-reverberant ratio (DRR), and clarity (C50) across 10 frequency bands using first-order Ambisonics (FOA) speech recordings as inputs. The proposed framework utilizes a novel feature named Spectro-Spatial Covariance Vector (SSCV), efficiently representing temporal, spectral as well as spatial information of the FOA signal. Our models significantly outperform existing single-channel methods with only spectral information, reducing estimation errors by more than half for all three acoustic parameters. Additionally, we introduce FOA-Conv3D, a novel back-end network for effectively utilising the SSCV feature with a 3D convolutional encoder. FOA-Conv3D outperforms the convolutional neural network (CNN) and recurrent convolutional neural network (CRNN) backends, achieving lower estimation errors and accounting for a higher proportion of variance (PoV) for all 3 acoustic parameters.

\end{abstract}


\section{Introduction}
In recent years, dynamic parameterization of acoustic environments has gained significant attention. Estimating acoustic parameters across different frequencies is crucial for simulating virtual sounds, ensuring Virtual Reality (VR) and Augmented Reality (AR) applications provide a congruent and plausible acoustic simulation~\cite{ar_vr}. Instead of measuring acoustic attributes from microphone captures using controlled signals such as sine sweeps~\cite{sin_sweep}, earlier studies focused on determining acoustic parameters from microphone captures of a prior unknown speech signal, known as blind acoustic parameters estimation~\cite{ace,blind_t60_cnn,auditory_inspired_acoustic,crnn_1,blstm_mask,joint_est_1}. However, these studies typically predict average acoustic parameters across all frequencies, failing to capture the natural frequency-dependent variations in acoustic parameters that happen across various spaces.

On the other hand, research into the estimation of acoustic contexts, such as Direction-of-Arrival (DoA) and reverberation time (T60), are increasingly relying on deep networks~\cite{joint_est_1,bryan2020impulse,doa_dnn_foa, doa_dnn_2,doa_dnn_3,doa_dnn_4,nofoa_deep_t60_summary,deep_T60_C50}. This shift is driven not only by their superior performance~\cite{bryan2020impulse,joint_est_1}, but also by the limitation of signal processing methods, which depend on specific environmental assumptions and struggle to generalize across different tasks~\cite{nofoa_trad_drr,nofoa_deep_t60_summary}. Existing deep learning based approaches are mainly developed to estimate acoustic parameters from a single-channel audio signal, which contains limited spatial information~\cite{cnn_encoder,crnn_1,joint_est_1,blstm_mask}. Recent research has shifted toward multi-channel signals. In~\cite{cnn_encoder}, the author proposed a Convolution Neural Network (CNN) based framework that extracted binaural features from dual-channel signals for T60 and room volume estimation. With multi-channel microphone rigs becoming more available, efficiently utilizing spatial information to enhance deep network estimation performance has emerged as a new challenge.

\begin{figure}[!ht]
\centering
\includegraphics[width=0.9\columnwidth]{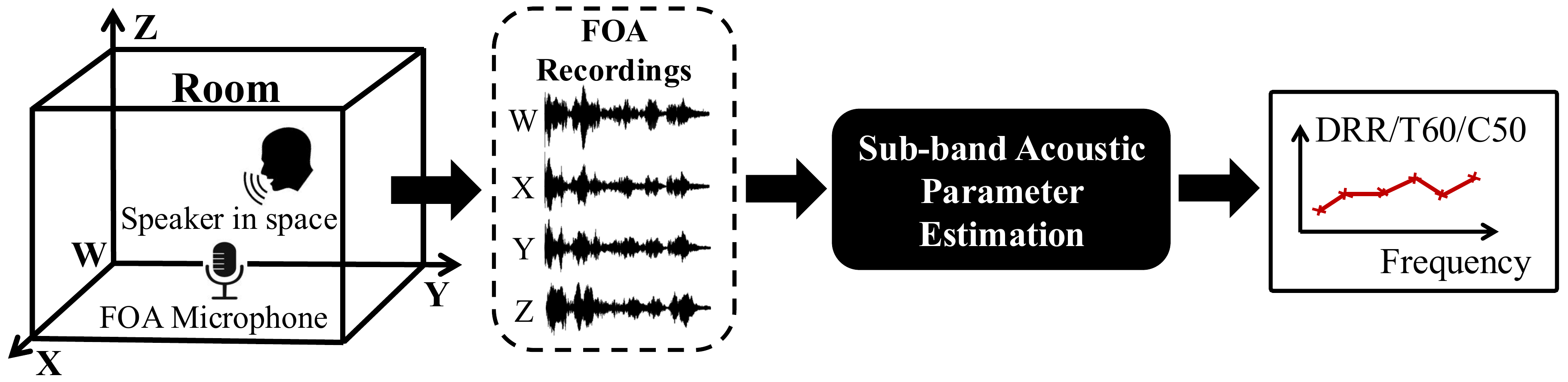} 
\caption{The problem context of this paper}
\label{intro_figure}
\end{figure}

First-order Ambisonics (FOA) is a prevalent audio recording format in AR and VR ~\cite{cambridge2021}, with its four channels (W, X, Y, Z) capturing a soundfield in all directions. The integration of FOA and deep networks has been well-established for DoA estimation ~\cite{doa_dnn_foa,hoa_doa_dnn, dcase19overiew}. However, its potential for estimating other acoustic parameters remains largely unexplored. Inspired by~\cite{cov_feature}, we propose a unified framework that derives the Spectro-Spatial Covariance Vector (SSCV) from FOA recordings to estimate frequency-varying DRR, T60, and C50, as illustrated in Figure~\ref{intro_figure}. To the best of our knowledge, this is the first method to utilize FOA inputs for blind estimation of frequency-dependent acoustic parameters.

\section{Frequency Varying Acoustic Parameters Estimation}
\label{sec:method}
\subsection{Spectro-Spatial Covariance Vector (SSCV)}
\begin{figure}[!ht]
\centering
\includegraphics[width=\columnwidth]{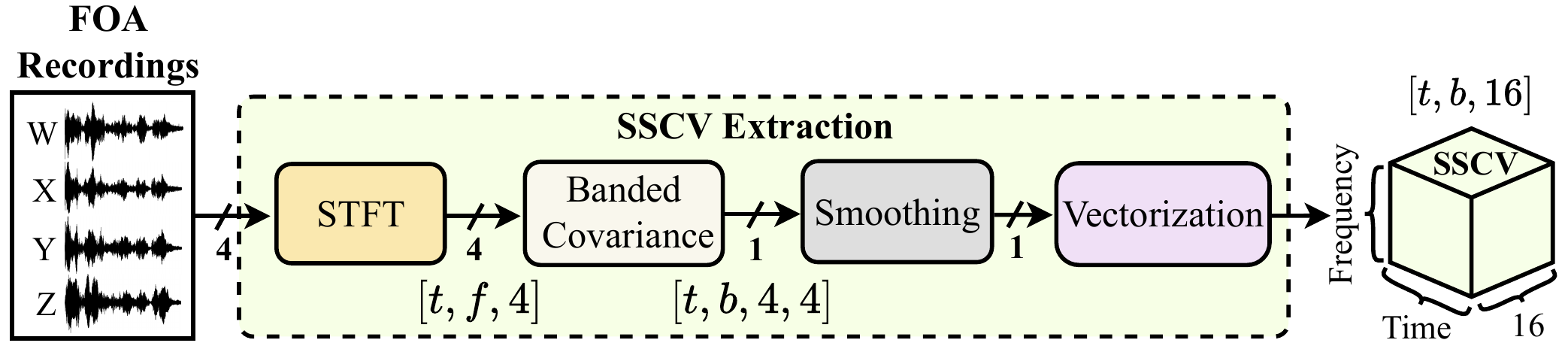} 
\caption{The process of SSCV feature extraction}
\label{sscv}
\end{figure}

The effective representation of multi-channel speech recordings has been extensively studied in tasks like target speaker extraction~\cite{meng24b_interspeech} and room volume estimation~\cite{cnn_encoder}. Prior research highlights the importance of incorporating time-frequency power distribution and inter-channel differences for robust performance. To address this need in FOA signals, we developed the SSCV feature, designed to encode both time-frequency power distribution and inter-channel correlation, including phase differences. This dual capability provides a comprehensive representation of spatial audio, crucial for acoustic parameter estimation. The extraction process, depicted in Figure~\ref{sscv}, involves three stages: banded covariance computation, smoothing, and vectorization.
\subsubsection{Banded Covariance}
Given an $M$-channel FOA signal \(\mathbf{x} = [x_1(t), x_2(t), \ldots , x_M(t)]\), where \(t=1, \ldots, N\), we aim to extract a covariance-based feature inspired from~\cite{cov_feature} that encapsulates the spatial characteristics of the sound field across time and frequency. We start by dividing the signals into windows of \(L\) samples with a hop size of \(H\), i.e.
\begin{equation}
\setlength{\abovedisplayskip}{3pt}
\setlength{\belowdisplayskip}{3pt}
 x_{i,n} =  x_i[nH:nH+L-1], \quad n \in \mathbb{N},
\end{equation}
where \(n\) is the frame index. Each frame is then transformed to the frequency domain using a Discrete Fourier Transform (DFT),
\begin{equation}
\setlength{\abovedisplayskip}{3pt}
\setlength{\belowdisplayskip}{3pt}
\mathbf{X}(n,f) = [\text{DFT}(x_{1,n}), \hdots, \text{DFT}(x_{M,n})]^T,
\end{equation}
where \(f\) refers to the DFT bin of the transformed data. We then compute a banded covariance matrix for each frame using a mel-filterbank to group the DFT bins into mel-scale bands. The banding operation can be expressed as
\begin{equation}
\setlength{\abovedisplayskip}{3pt}
\setlength{\belowdisplayskip}{3pt}
\text{Cov}_{\mathbf{x}}(n,b) = \frac{1}{|\mathcal{B}_b|} \sum_{f' \in \mathcal{B}_b} W_b(f')\cdot \mathbf{X}(n,f') \cdot \mathbf{X}(n,f')^H,
\end{equation}
where \(b\) is the band index for the mel-filterbank, $W_b(f')$ is the filter weight at band $b$ and DFT bin $f'$, and $\mathcal{B}_b$ is the set of DFT bin indices included in the \(b\)-th band. We operate on the assumption that $\sum_{f'}W_b(f')=1$.
\subsubsection{Smoothing}
We apply a one-pole smoothing filter to the banded covariance data, expressed via
\begin{equation}
    \text{Cov}'_{\mathbf{x}}(n,b) = (1 - \alpha) \cdot \text{Cov}_{\mathbf{x}}(n,b) + \alpha \cdot \text{Cov}'_{\mathbf{x}}(n-1,b),
\end{equation}
where \(\alpha\) is a learnable smoothing factor optimized jointly with the back-end, enabling the system to infer the optimal level of temporal smoothing in each band.
\subsubsection{Vectorization}
For each sub-band and frame, the resulting smoothed covariances are complex-valued \(M \times M\) Hermitian matrices. The diagonal elements give the energy of each channel, while off-diagonal conjugate pairs encode correlation and (average) phase differences between channels. The final step in extracting the SSCV involves vectorizing these matrices into real-valued vectors with \(M^2\) elements. 

We first construct an initial column vector $\mathcal{C}_{n,b}$ by sorting the diagonal and off-diagonal elements of the smoothed covariance \(\text{Cov}'_{\mathbf{x}}(n,b)\). The diagonal entries are placed in order in the first \(M\) elements of $\mathcal{C}_{n,b}$, while the remaining \(M \cdot (M-1)\) elements contain the \(\frac{M \cdot (M-1)}{2}\) unique conjugate pairs, where ordering of the pairs can be arbitrary.

$\mathcal{C}_{n,b}$ is then converted to a real-valued vector, $\mathcal{R}_{n,b}$, by transforming the diagonal entries and off-diagonal pairs separately. For the diagonal entries, we apply a real DFT matrix \(\mathbf{F}\) of size \(M \times M\)~\cite{real_DFT}, such that $\mathcal{R}_{n,b}[1:M] = \mathbf{F} \mathcal{C}_{n,b}[1:M]$. For each subsequent pair of off-diagonal entries, we apply the transformation,
\begin{equation}
    \mathcal{R}_{n,b}[i:i+1] = \begin{bmatrix} 1/\sqrt{2} & 1/\sqrt{2} \\ -i/\sqrt{2} & i/\sqrt{2} \end{bmatrix} \mathcal{C}_{n,b}[i:i+1]
\end{equation}
where $i = M+1,M+3,\hdots,M^2-1$. This transformation decouples the real and imaginary parts of each conjugate pair.

Finally, we take the logarithm of the first element and normalize the entire vector by this first element. The SSCV for the \(n\)-th frame and \(b\)-th band is given by:
\begin{equation}
\setlength{\abovedisplayskip}{3pt}
\setlength{\belowdisplayskip}{3pt}
    \text{SSCV}(n,b) = \left[ \log(\mathcal{R}_{n,b}[0]), \frac{\mathcal{R}_{n,b}[1]}{\mathcal{R}_{n,b}[0]}, \ldots, \frac{\mathcal{R}_{n,b}[M^2-1]}{\mathcal{R}_{n,b}[0]} \right].
\end{equation}

\begin{figure*}[!ht]
\centering
  \begin{minipage}{0.5\textwidth}
    \centering
    \vspace{-2mm}
    \subfigure[FOA-CNN Model]{
      \includegraphics[width=0.95\textwidth]{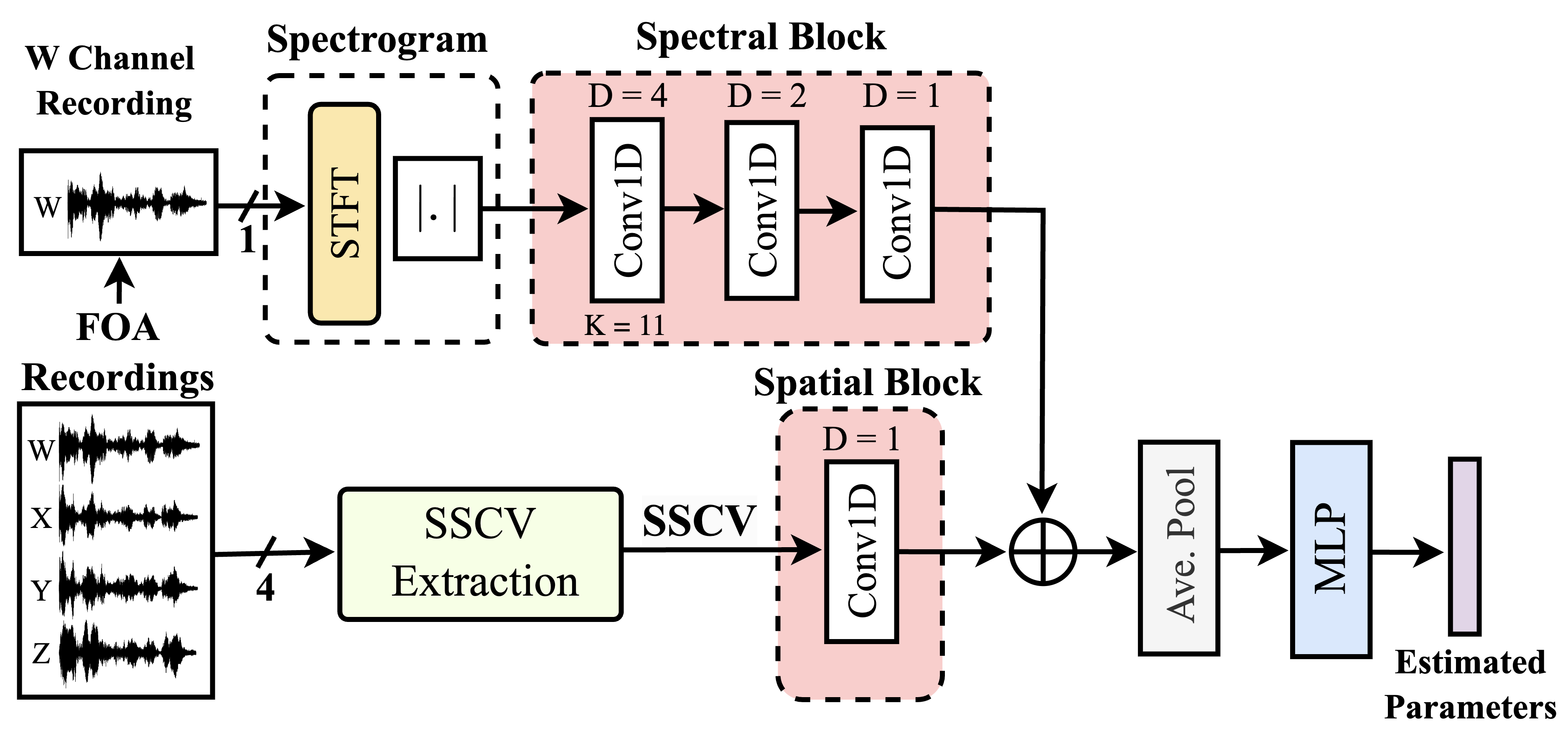}\label{fig:network_a}}
  \end{minipage}%
  \hfill
  \begin{minipage}{0.5\textwidth}
    \centering
    \vspace{-2mm}
    \subfigure[FOA-CRNN Model]{
      \includegraphics[width=0.9\textwidth]{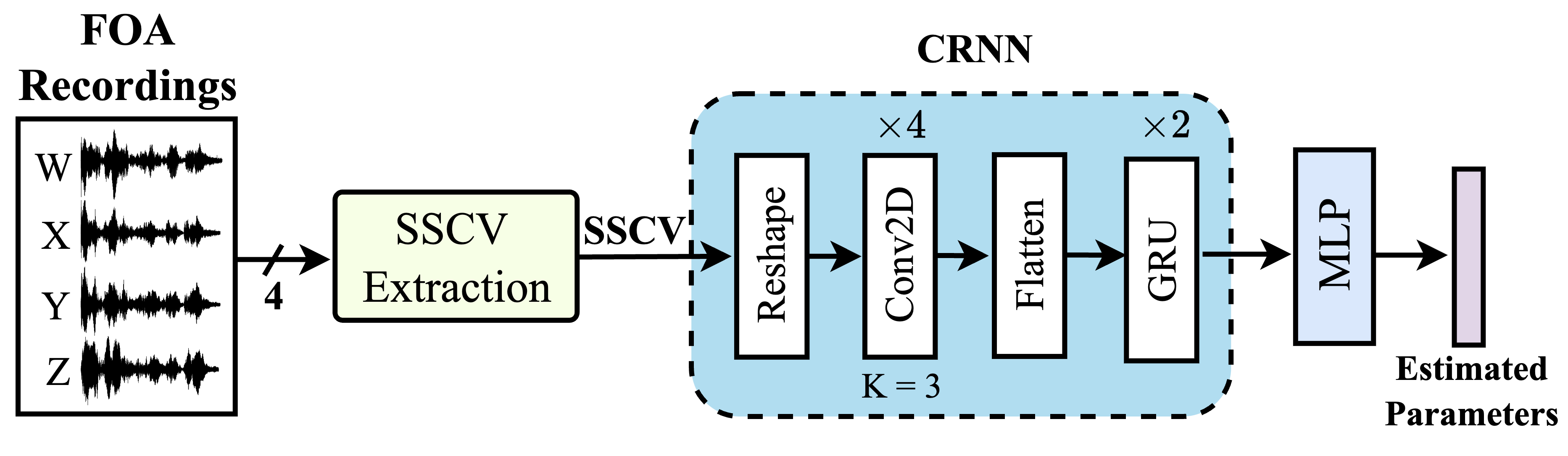}\label{fig:network_b}}\\
    \vspace{-2mm}
    \subfigure[FOA-Conv3D Model]{
      \includegraphics[width=0.9\textwidth]{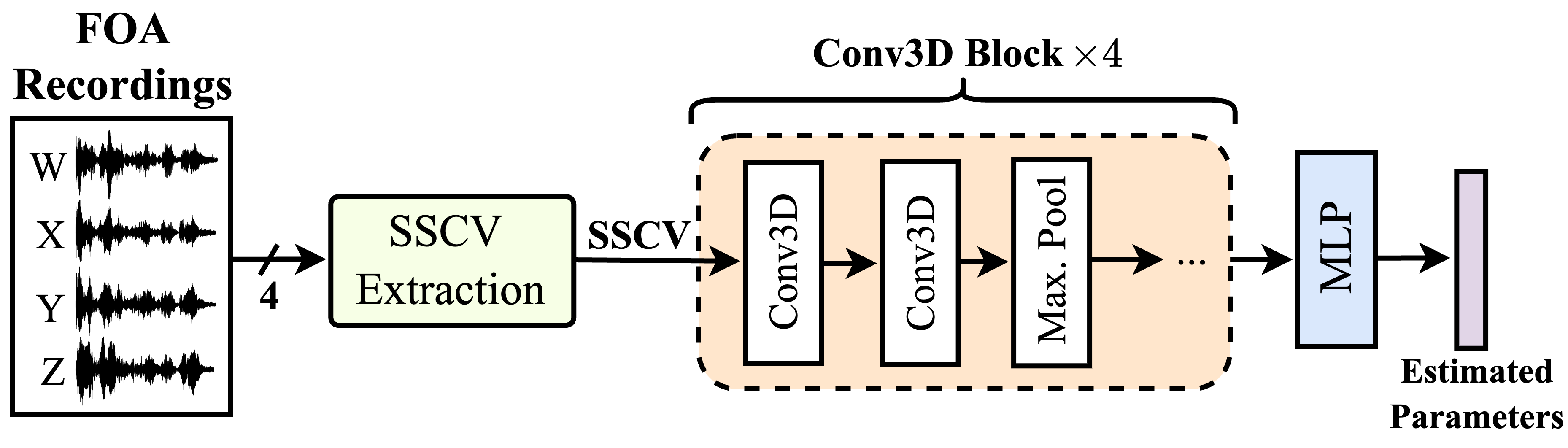}\label{fig:network_c}}
  \end{minipage}
  \vspace{-3.5mm}
  \caption{An Overview of the FOA based network structures applied in this paper (K:Kernel size, D:Dilation size)}
  \label{fig:model_overview}
  \vspace{-4mm}
\end{figure*}
\subsection{Back-end Model Architectures}
\label{model_architecture}
The back-end model consists of two main components: a feature encoder and a Multi-Layer Perceptron (MLP). The feature encoder transforms computed features into an embedding, known as the room fingerprint~\cite{reverb_fingerprint}. We employed three encoder structures: CNN, CRNN, and Conv3D, as shown in Figure~\ref{fig:model_overview}. The CNN and CRNN architectures are adapted from existing models~\cite{crnn_1, cnn_encoder}.

The FOA-CNN model (Figure~\ref{fig:network_a}) includes two branches with spectral blocks comprising three separable convolutions interspersed with ReLU activations and layer normalization~\cite{layer_norm}. The FOA-CRNN model (Figure~\ref{fig:network_b}) features four 2D-convolutional layers followed by max-pooling, ELU activation, and two GRUs, one bidirectional and one unidirectional.

We also propose a novel FOA-Conv3D structure (Figure~\ref{fig:network_c}), inspired by~\cite{spatial_librispeech}. This model inputs the SSCV into four Conv3D blocks with 32, 64, 128, and 256 channels, each containing two 3D-convolutional layers with a $1\times3\times3$ kernel. The layers capture inter-channel power and phase differences, with the first layer matching input dimensions and the second expanding channels. A 3D max-pooling layer reduces time and frequency dimensions.

The encoded outputs from all encoders are flattened and passed through an MLP with three fully-connected layers to produce parameter value estimates for 10 sub-bands.

\section{Experimental Setups}
\label{sec:experiment}
\subsection{Dataset}
We used the Spatial Librispeech-Lite dataset~\cite{spatial_librispeech} for our experiments.
Since no official validation set is provided, we uniformly sampled 1,000 utterances from the training set, resulting in 16,195 training, 1,000 validation, and 9,901 test utterances. All utterances were either trimmed or padded to a duration of 4 seconds, with a 16 kHz sampling rate. The set contains the simulated FOA recordings by convolving the synthetic room impulse responses (RIRs) with randomly assigned LibriSpeech samples. The training set includes FOA recordings in 8,952 simulated rooms and there are 4,970 rooms for the test set. There are 20 geometrical configurations for each room and the sound sources are randomly assigned to each combination of room and geometrical configuration as illustrated in~\cite{spatial_librispeech}. 

The dataset provides acoustic parameter labels in third-octave bands derived from RIRs. In this study, we focused on the DRR, T60, and C50 within the frequency range of 1 kHz to the Nyquist frequency (8 kHz), containing 10 sub-bands (\{1, 1.25, 1.6, 2, 2.5, 3.15, 4, 5, 6.3, 8\}kHz). We excluded room parameters below 1kHz to reduce the potential effect of room modes and to ensure that the source speech signals contain sufficient energy in the frequency bands of interest. We represent reverberation time as T60 by doubling the provided T30 values.

\subsection{Model Configuration}
As described in Section~\ref{model_architecture}, we applied three encoder structures: FOA-CNN, FOA-CRNN, and FOA-Conv3D, along with two baseline single-channel models: Single-CNN~\cite{cnn_encoder} and Single-CRNN~\cite{crnn_1}. The W-channel from the FOA recordings was used as the single-channel signal. For the Single-CNN, the model consisted of solely the (mono) upper branch of Figure~\ref{fig:network_a}. The short-time Fourier transform (STFT) employed 96ms Hann windows with 50\% overlap. Single-CRNN used 20 Mel Frequency Cepstral Coefficients (MFCC) as inputs, computed with a 25ms frame size and 10ms steps. The segmentation and windowing for the SSCV are chosen to be identical to the STFT in Single-CNN in order to align frames for single-channel and FOA-based models. We chose the Mel-filterbank with 52 triangular filters following the mel-scale for frequency banding. To ensure fair model comparisons, we used the same SSCV settings for the FOA-CRNN and FOA-Conv3D models.

\subsection{Training and Evaluation}
All models in this study were initially set to train for 100 epochs using the Adam optimizer~\cite{adam} with an initial learning rate of 0.0005 and a $20\%$ dropout rate applied to the output of the convolutional block. The learning rate was halved if the validation loss did not improve for five consecutive epochs, and early stopping was employed if the validation loss did not decrease for 10 epochs. The batch size was set to 128. For each parameter, optimization was performed using the mean square error (MSE) for all sub-bands as the loss function. For the T60 estimation in our proposed model, the logarithm of both the reference and estimated T60 values were used when calculating the MSE, as psychoacoustic studies indicate that humans perceive T60 on a logarithmic scale~\cite{log_rt60}. The model with the lowest validation loss was selected for evaluation.

Performance evaluation employed three primary metrics: Mean Absolute Error (MAE), Proportion of Variance (PoV), and Pearson Correlation Coefficient (PCC). We included MAE and PCC as they are two commonly used metrics in this task~\cite{berp,blstm_mask,crnn_1,phase_info}, but we also introduce the PoV metric as it can effectively quantify the extent to which the model captures the variability of the data~\cite{pov_book}. Given that different acoustic parameters have varying value ranges, we believe PoV can serve as a more informative metric for evaluating the reliability of a unified model in this problem context. 

\section{Results and Discussion}
\label{sec:results}
\begin{figure*}[!ht]
\centering
  \includegraphics[width=0.88\textwidth]{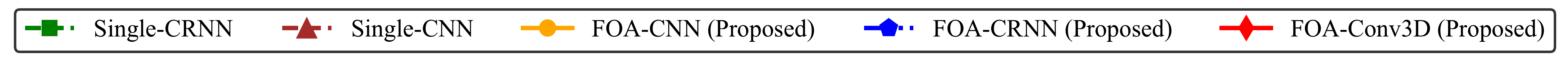}
  \vspace{-0.1cm}
  \subfigure[MAE for DRR Estimation]{
\includegraphics[width=.32\textwidth]{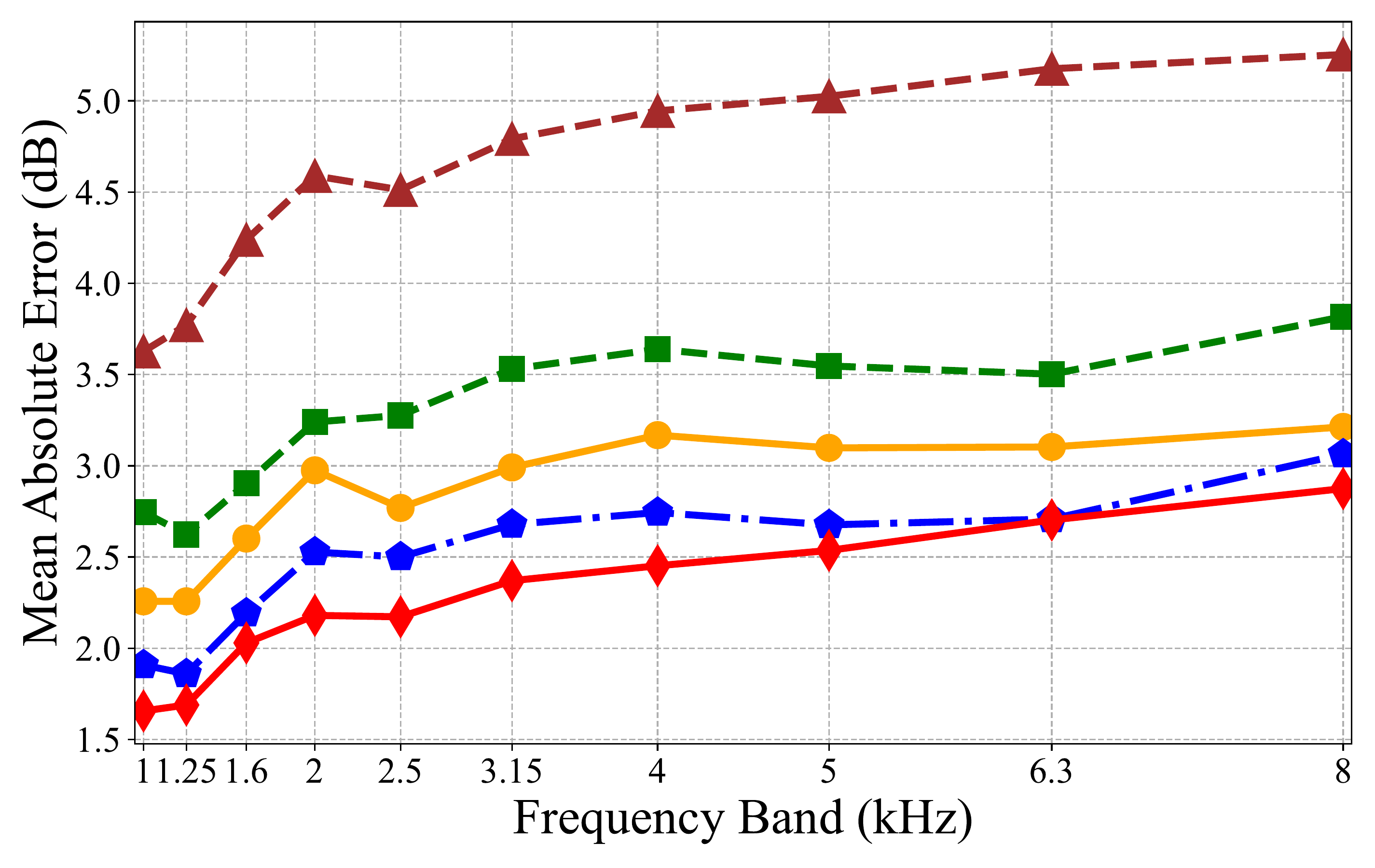}\label{fig:drr_mae}}
    \hfill
  \subfigure[MAE for T60 Estimation]{
    \includegraphics[width=.32\textwidth]{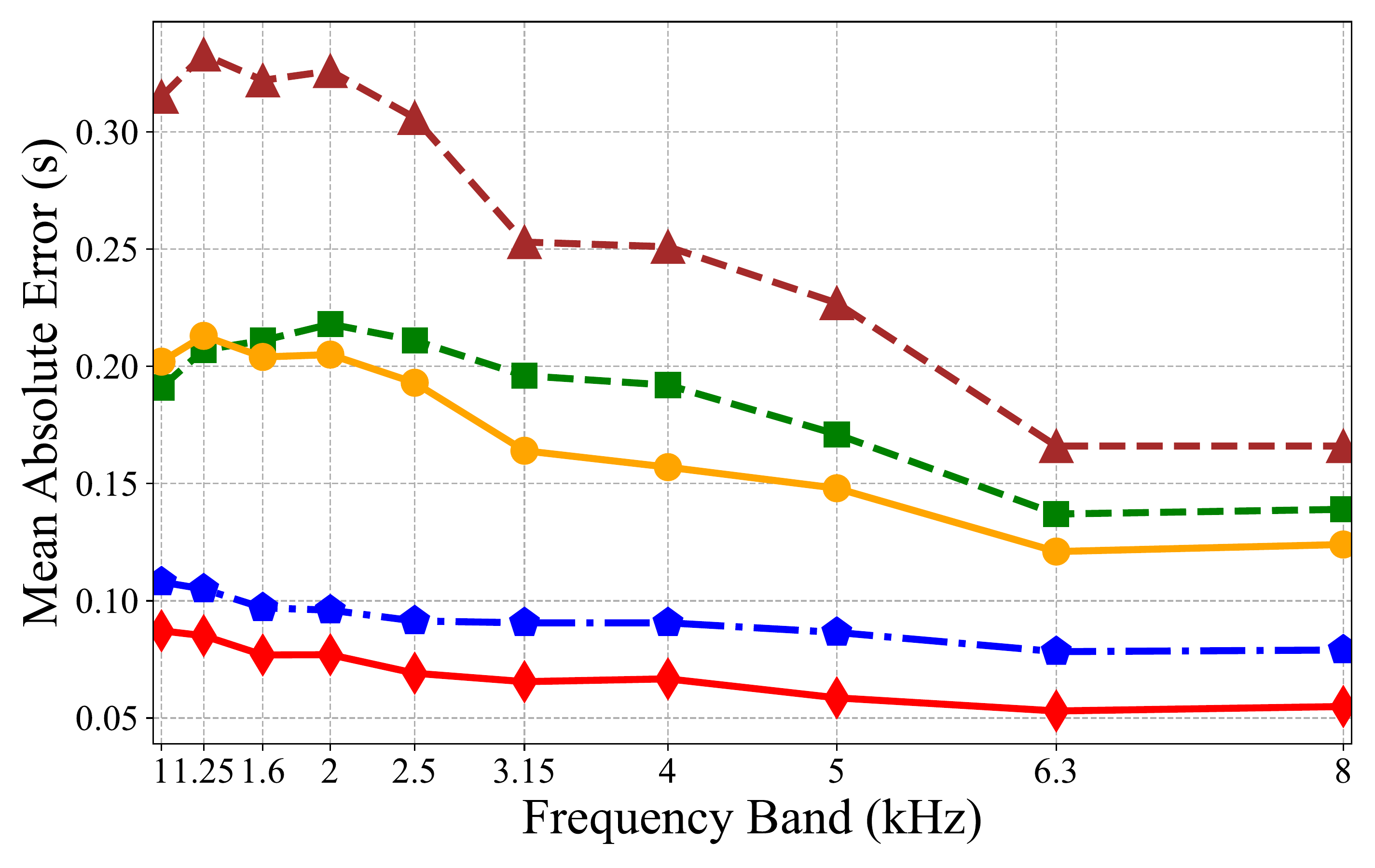}\label{fig:t60_mae}}
    \hfill
  \subfigure[MAE for C50 Estimation]{%
    \includegraphics[width=.32\textwidth]{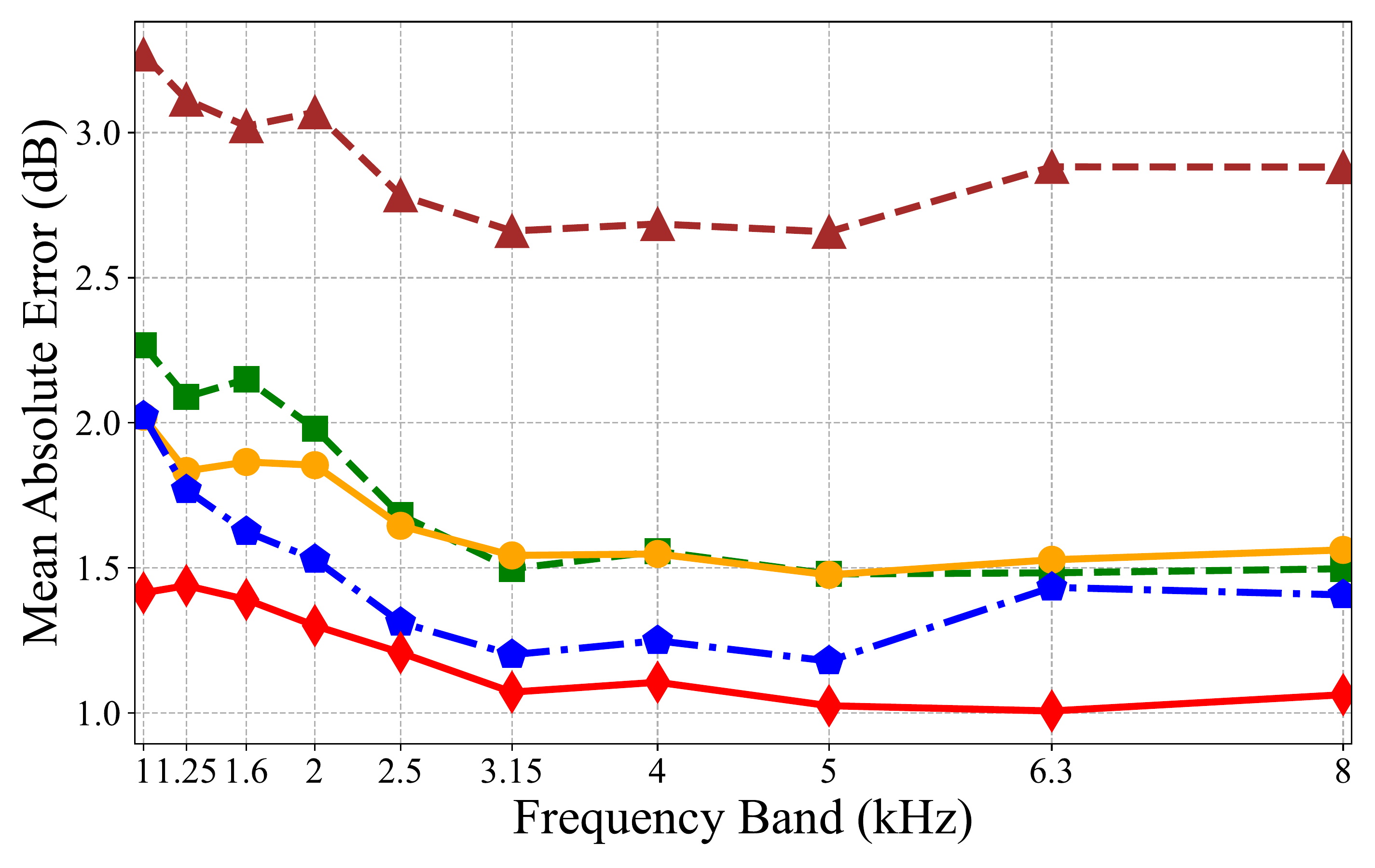}\label{fig:c50_mae}}
    \hfill
  \subfigure[PoV for DRR Estimation]{%
    \includegraphics[width=.32\textwidth]{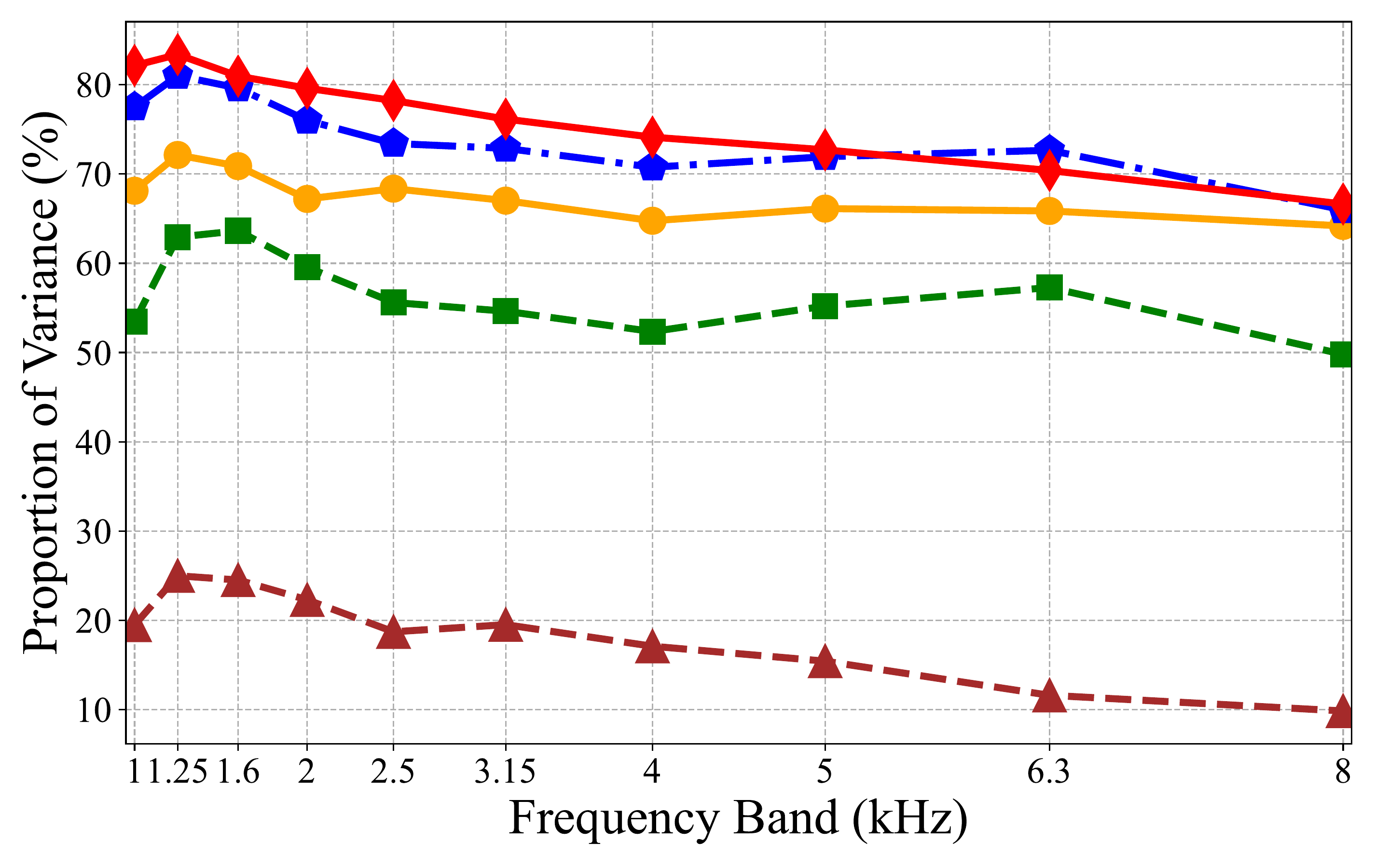}\label{fig:drr_pov}}
    \hfill
    \hspace{-0.1cm}
  \subfigure[PoV for T60 Estimation]{%
    \includegraphics[width=.32\textwidth]{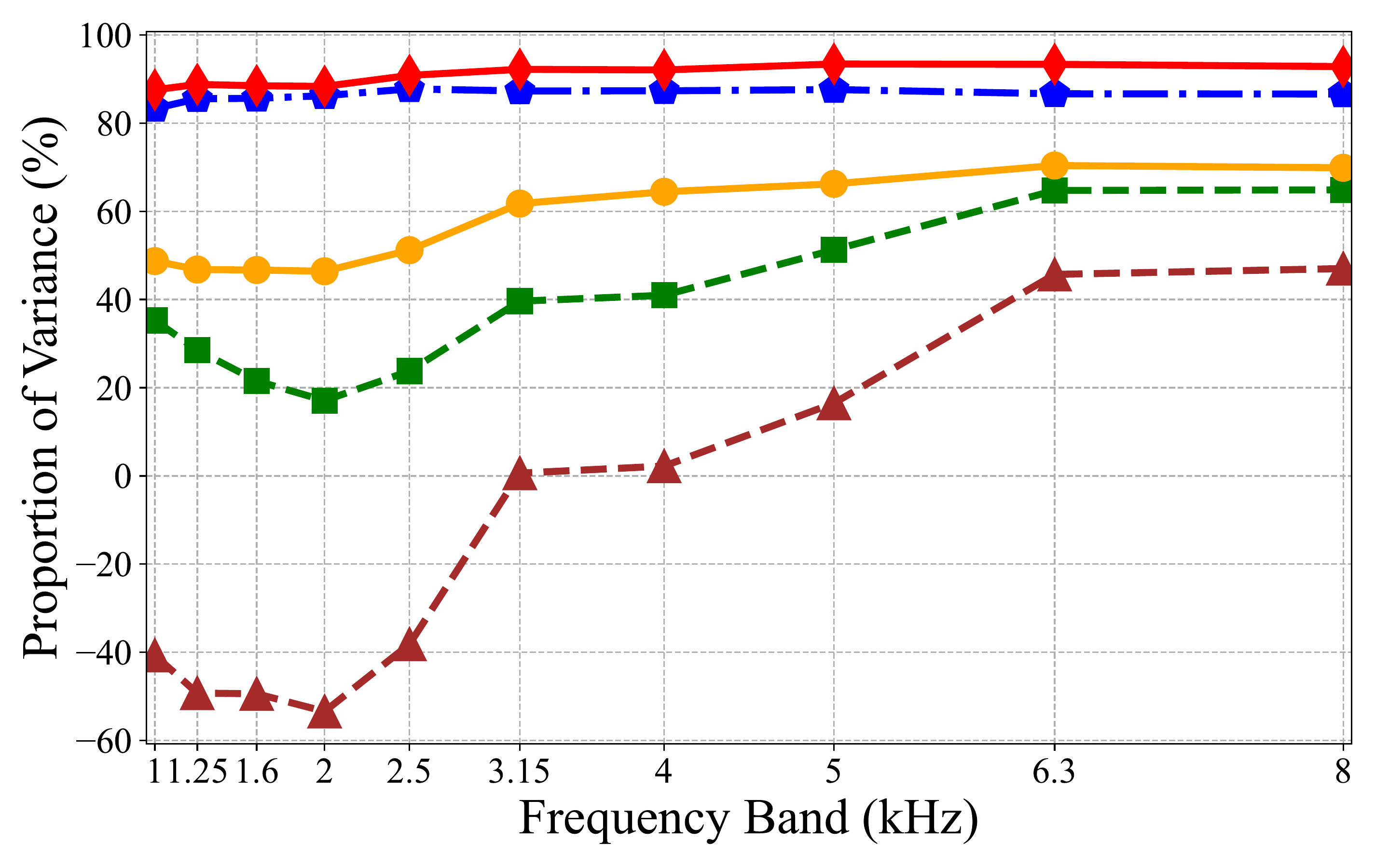}\label{fig:t60_pov}}
    \hfill
    \hspace{-0.1cm}
  \subfigure[PoV for C50 Estimation]{%
    \includegraphics[width=.32\textwidth]{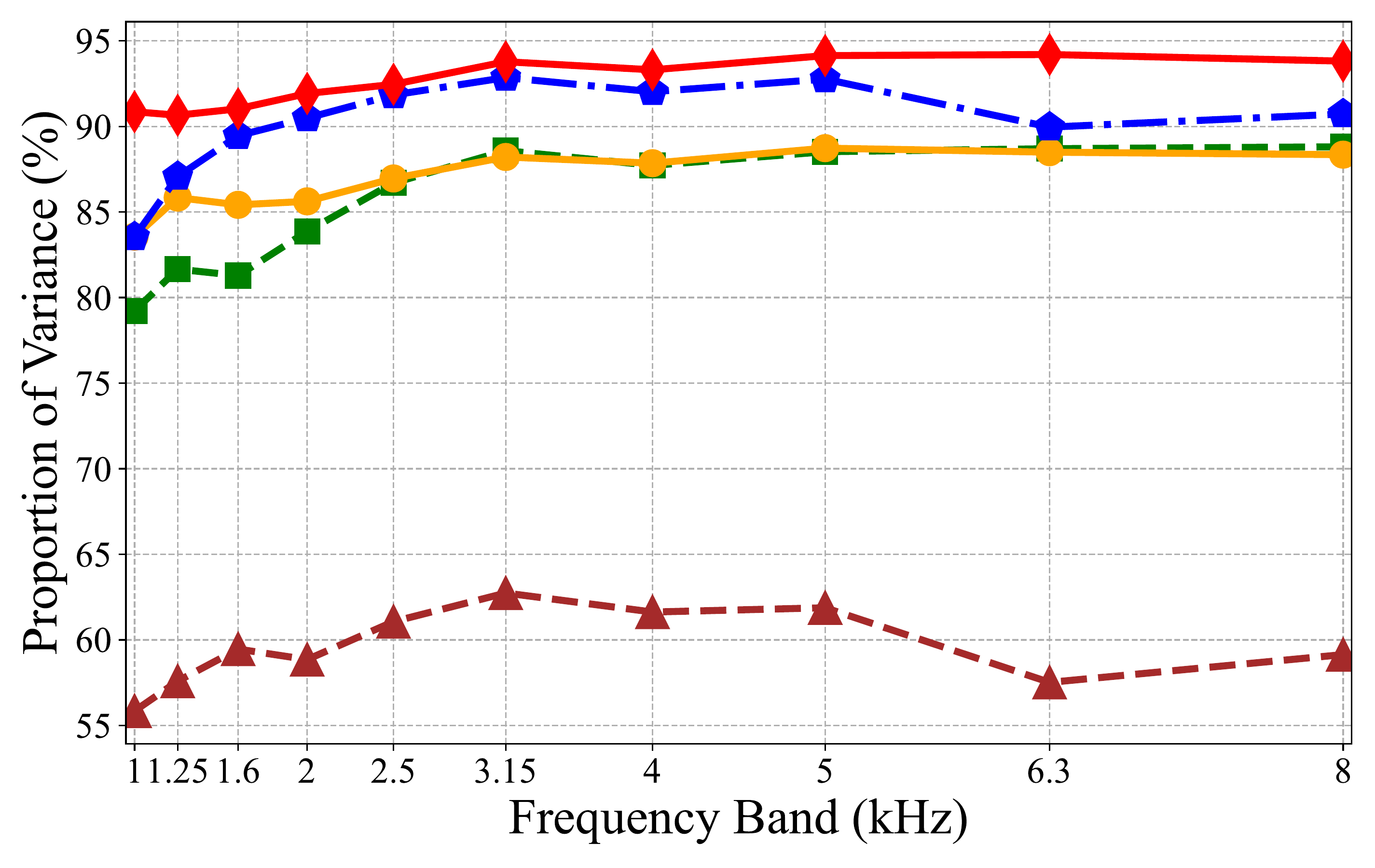}\label{fig:c50_pov}}
    \hfill
  \subfigure[PCC for DRR Estimation]{%
    \includegraphics[width=.32\textwidth]{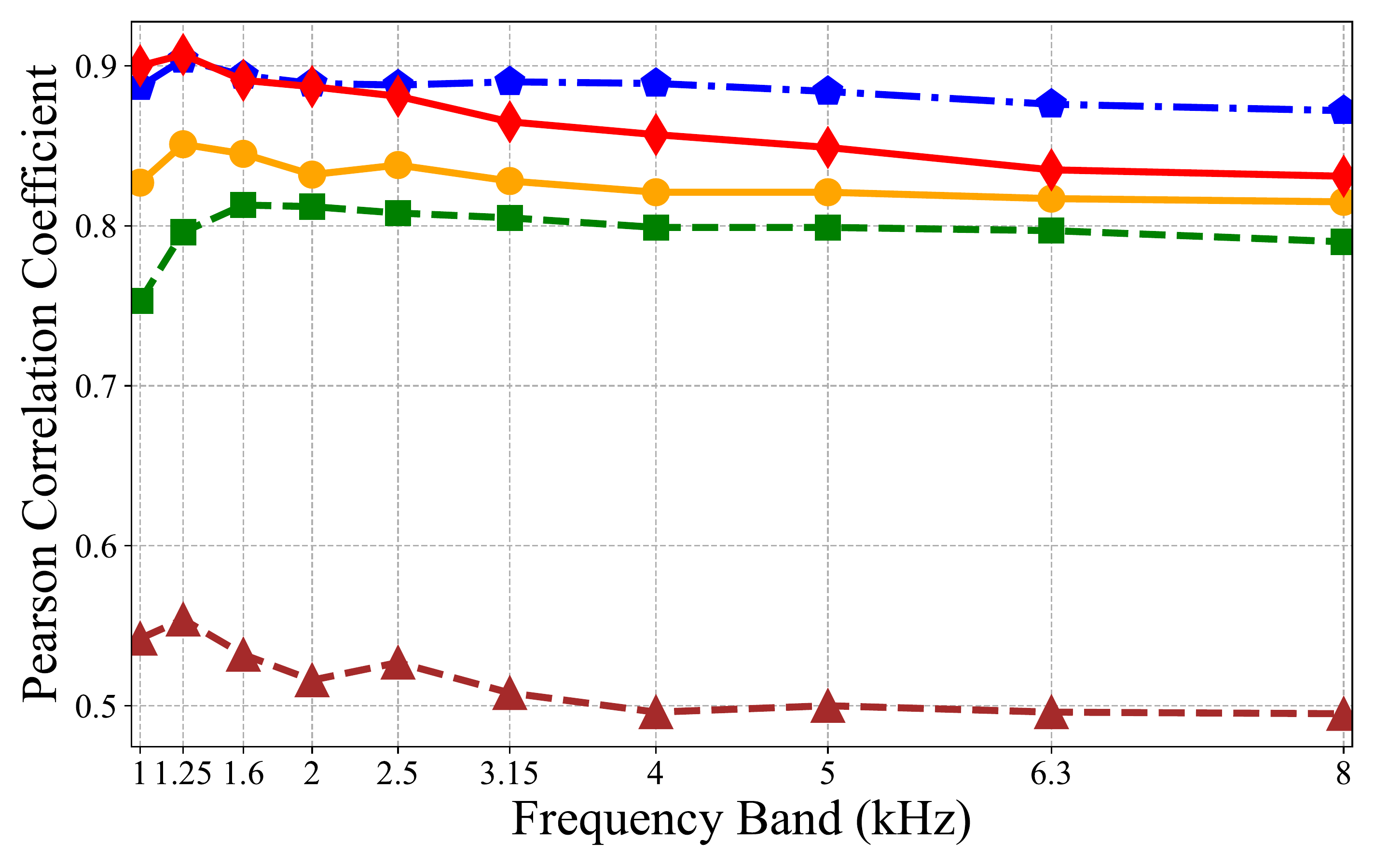}\label{fig:drr_pcc}}
    \hfill
  \subfigure[PCC for T60 Estimation]{%
    \includegraphics[width=.32\textwidth]{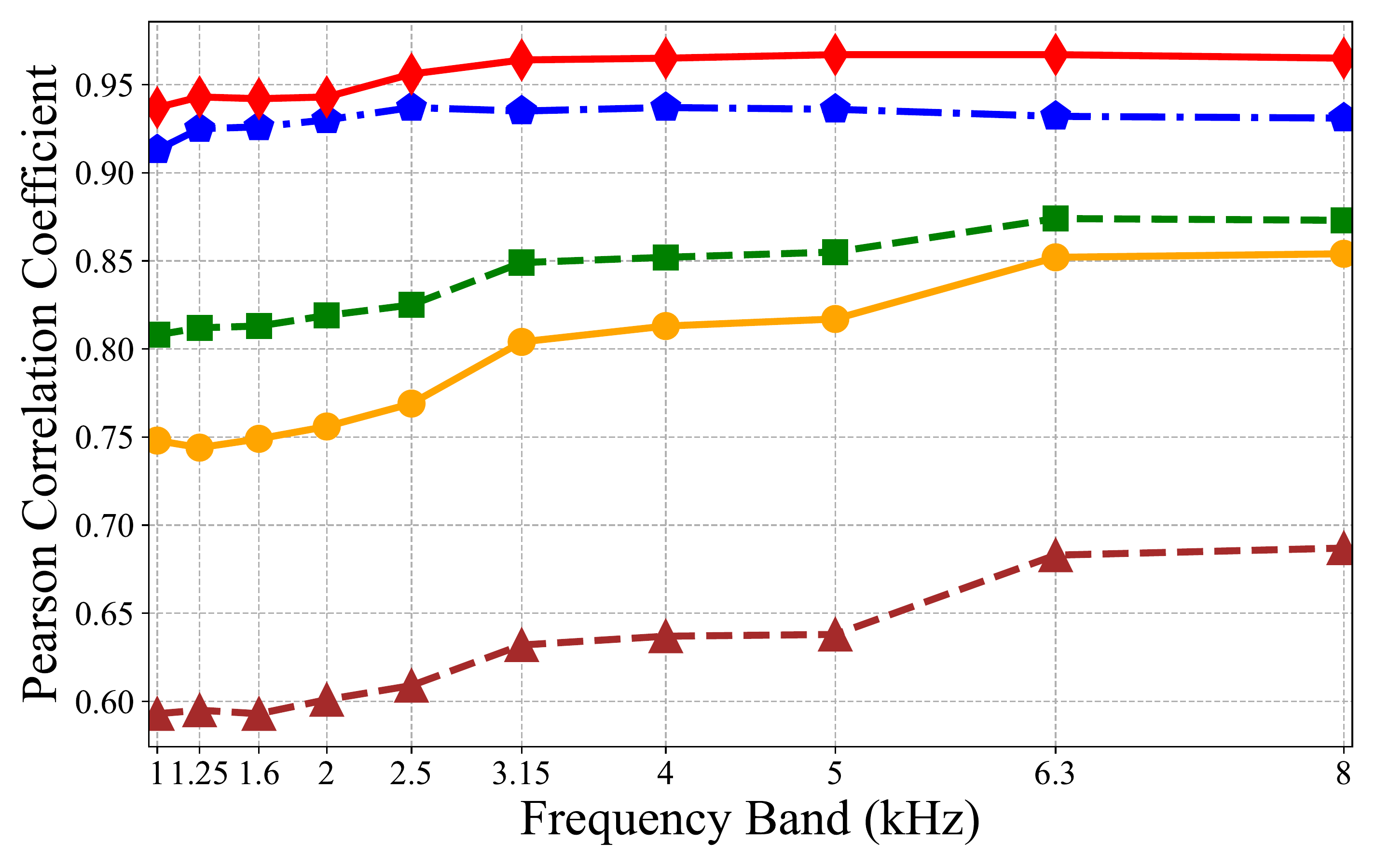}\label{fig:t60_pcc}}
    \hfill
  \subfigure[PCC for C50 Estimation]{%
    \includegraphics[width=.32\textwidth]{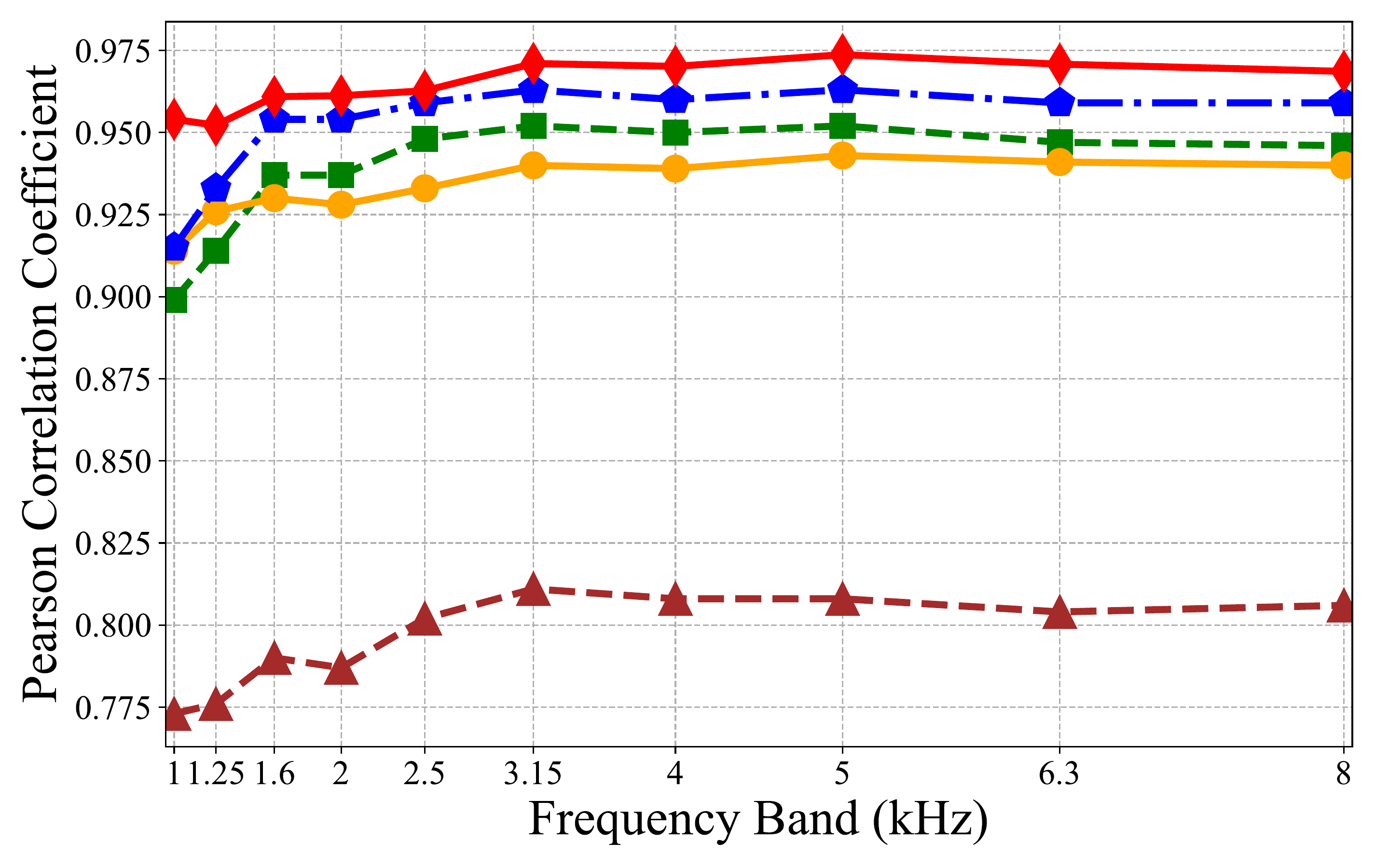}\label{fig:c50_pcc}}
  \vspace{-2mm}
  \caption{Model performance metrics (MAE $\downarrow$, PoV $\uparrow$, PCC $\uparrow$; ordinate) for all three acoustic parameter estimation tasks (DRR, T60, C50) as a function of frequency (abscissa). Different curves represent different models (see legend). } 
  \label{fig:results}
\end{figure*}

The MAE, PoV, and PCC for each model and each parameter considered in this study are presented in Figure~\ref{fig:results}.
\subsection{Benefit of adding spatial information}
When comparing the single-channel and FOA versions of otherwise similar frameworks, such as the Single CNN and FOA-CNN, we observe a significant performance improvement across all frequency bands and all 3 performance metrics for FOA input. A similar, significant performance improvement is observed for the CRNN network. These improvements demonstrate the benefit of capturing spatial information for acoustic parameter estimation tasks. More importantly, it demonstrates the effectiveness of SSCV to describe such information from FOA audio captures, and subsequently allows the neural network to more accurately predict the room acoustic parameters.

\subsection{Benefit of recurrent model structures}
From the results presented in Figure~\ref{fig:results}, it is evident that the CNN-encoder performs the worst compared to other model architectures under the same input configurations (Single or FOA). This result could be attributed to the average pooling operation after the embedding concatenation. Averaging across the time dimension eliminates crucial temporal information, which seems beneficial for accurate acoustic parameter estimation. We hypothesize that model structures that have access to  spectral and temporal information such as CRNN are more capable of separating temporally-distinct events such as the direct sound, early reflections and late reverberation~\cite{temporal_infos_cnn}. 

Our proposed FOA-Conv3D back-end structures takes this approach one step further by encoding time, frequency and inter-channel information by means of employing multiple Conv3D blocks and SSCV features as input. With the exception of the PCC performance metric for DRR estimation, the FOA-Conv3D model outperforms all other models, indicating that learning information across three dimensions (temporal, spectral and spatial) and representing that information by means of the SSCV features is very effective to predict acoustic room attributes from FOA microphone captures.

\section{Conclusion}
In this study, we developed a unified framework for blind estimation of DRR, T60, and C50 across 10 frequency bands from FOA recordings, introducing SSCV as a novel spatial audio representation. Experimental results indicate that our model significantly improves deep learning-based acoustic parameters estimation compared to existing single-channel approaches. Moreover, the proposed FOA-Conv3D model, which can utilize the SSCV features across time and frequency effectively, delivers state-of-the-art estimation performance compared to earlier approaches based on mono input and/or alternate model architectures. Future work will extend this framework to other acoustic context attributes, such as room geometry and source orientation, and explore alternative back-end architectures which can be integrated with our SSCV feature encoder.

\bibliographystyle{IEEEbib}
\clearpage
\bibliography{mybib}
\vspace{12pt}
\end{document}